\documentclass[aps,pra,preprint,a4paper,showpacs,showkeys,superscriptaddress,nofootinbib]{revtex4-1}

\usepackage[utf8]{inputenc}
\usepackage[T1]{fontenc}
\usepackage{latexsym}
\usepackage{amsmath}
\usepackage{amssymb}
\usepackage{graphicx}
\usepackage{subfigure}
\usepackage{xcolor}
\usepackage{cancel}
\usepackage{physics}
\usepackage{subfiles}
\usepackage{multirow,tabularx,boldline,array}
\usepackage{hyperref}     
\usepackage[capitalise]{cleveref}
\crefformat{equation}{(#2#1#3)}
\crefrangeformat{equation}{(#3#1#4) - (#5#2#6)}
\crefmultiformat{equation}{(#2#1#3)}{ - (#2#1#3)}{, (#2#1#3)}{ - (#2#1#3)}

\newcolumntype{C}[1]{>{\centering\arraybackslash}p{#1}}
\hypersetup{colorlinks,%
  citecolor=blue,%
  linkcolor=cyan}%
\usepackage[titletoc]{appendix}
\usepackage{enumerate}
\DeclareMathOperator{\ext}{ext}

\graphicspath{{./figures/}}
\begin{document}


\title{Entanglement entropy of asymptotically flat
non-extremal and extremal black holes with an island}

\author{Wontae Kim}%
\email[]{wtkim@sogang.ac.kr}%
\affiliation{Department of Physics, Sogang University, Seoul, 04107,
	Republic of Korea}%
\affiliation{Center for Quantum Spacetime, Sogang University, Seoul 04107, Republic of Korea}%

\author{Mungon Nam}%
\email[]{clrchr0909@sogang.ac.kr}%
\affiliation{Department of Physics, Sogang University, Seoul, 04107,
  Republic of Korea}%
\affiliation{Center for Quantum Spacetime, Sogang University, Seoul 04107, Republic of Korea}%

\date{\today}

\begin{abstract}
The island rule for the entanglement entropy is applied to an
eternal Reissner-Nordstr\"om black hole.
The key ingredient is that the black hole is assumed to be in thermal equilibrium with a heat bath
of an arbitrary temperature and so the generalized entropy is treated as being off-shell.
Taking the on-shell condition to the off-shell generalized entropy,
we find the generalized entropy and then obtain the entanglement entropy following the island rule.
For the non-extremal black hole, the entanglement entropy grows linearly in time
and can be saturated after the Page time as expected. The entanglement entropy also has a well-defined Schwarzschild limit.
In the extremal black hole, the island prescription provides a logarithmically growing entanglement entropy in time
and a constant entanglement entropy after the Page time.
In the extremal black hole, the boundary of the island
hits the curvature singularity where the semi-classical approximations appear invalid.
To avoid encountering the curvature singularity,
we apply this procedure to the Hayward black hole regular at the origin.
Consequently, the presence of the island in extremal black holes can provide a finite entanglement entropy, which
might imply non-trivial vacuum configurations of extremal black holes.

\end{abstract}

%


\keywords{Hawking radiation, Entanglement entropy, Extremal black holes, Page curve, Island}

\maketitle


\newpage
\section{introduction}
\label{sec:introduction}

Since Hawking's discovery that black hole radiates
\cite{Hawking:1974sw},
the information loss paradox has been one of the most intriguing issues in quantum gravity \cite{Hawking:1976ra}.
If unitarity in quantum mechanics is advocated, then information of black holes
should be conserved during the process of evaporation, of course,
the word ``unitarity'' can be used in different
meaning \cite{Unruh:2017uaw}.
From the viewpoint of the entanglement entropy of thermal radiation in evaporating black holes, it starts with the zero entropy and reaches a peak at the Page time and finally ends up with
the zero entropy \cite{Page:1993wv}. After the Page time, the entanglement entropy of radiation
follows the curve of the decreasing thermal entropy of black holes.

On the other hand,
a holographic derivation of the entanglement entropy was proposed
from AdS/CFT correspondence \cite{Ryu:2006bv,Hubeny:2007xt}, and quantum corrections to the entanglement entropy
were also studied in
Refs. \cite{Barrella:2013wja,Faulkner:2013ana}.
Then, the entanglement entropy was calculated at arbitrary orders in the bulk Planck constant
by using the quantum extremal surface which extremizes the generalized entropy
of the sum of area and bulk entanglement entropy \cite{Engelhardt:2014gca}.
In general, a density matrix at a region outside of black hole is
obtained by a partial trace over the state in complementary region.
The states inside a surface so called the island region \cite{Penington:2019npb,Almheiri:2019psf,Almheiri:2019yqk,Almheiri:2019hni}
need to be removed from the states which are traced out.
The emergence of islands was also shown by using the replica trick \cite{Penington:2019kki,Almheiri:2019qdq}.
Thus, the entanglement entropy at some region of Hawking radiation \( R \) should be represented
by the union of \( R \) and its island region \( I \). According to the prescription of minimal quantum extremal surface \cite{Engelhardt:2014gca},
the fine-grained entanglement entropy of the Hawking radiation is proposed as \cite{Almheiri:2019hni}
\begin{equation}\label{eq:generalized entropy}
	S=\min{\left\{  \ext S_{\rm gen} \right\}}= \min\left\{\ext\left[\frac{\mathcal{A}(\partial I)}{4G_N} + S_{\rm matter}(R\cup I)\right]\right\},
\end{equation}
where $S_{\rm gen}$ is the
generalized entropy and \( \mathcal{A}(\partial I) \) is the area of the boundary of island \( I \).
The matter part of the entanglement entropy in Eq. \eqref{eq:generalized entropy}
includes the UV divergence proportional to the island area, subject to a short distance proper cut-off scale
\cite{Bombelli:1986rw,Srednicki:1993im}, and thus,
the Newton constant \( G_N \) must be renormalized \cite{Susskind:1994sm}.
The prescription of minimal quantum extremal surface was initially applied to the two-dimensional AdS spacetime, and then
it was applied to arbitrary dimensions \cite{Almheiri:2019psy} and an
eternal Schwarzschild black hole \cite{Hashimoto:2020cas} as well as various interesting black holes \cite{Anegawa:2020ezn,Bak:2020enw,Ling:2020laa,Karananas:2020fwx,Wang:2021woy} with discussions on important role of gravitating bath
\cite{Krishnan:2020oun,Almheiri:2020cfm,Krishnan:2020fer,Geng:2021wcq,Ghosh:2021axl}.

In particular, for an extremal Reissner-Nordstr\"om black hole,
the entropy was shown to vanish for an arbitrary temperature of a heat bath in the Euclidean action formalism \cite{Hawking:1994ii}.
In addition, it was proposed that the entropy can vanish
when taking a vanishing temperature by employing 't Hooft's approach \cite{tHooft:1984kcu}
evaluating black hole entropy through a statistical-mechanical counting of states for a scalar field propagating outside the event horizon \cite{Demers:1995dq}.
Of course, a non-zero entropy was also derived for a stringy extremal black hole by microscopic counting of degrees of freedom in sting theory
without resort to mentioning the status of the temperature \cite{Strominger:1996sh}.
On the other hand, the entropy of extremal black holes turned out to be
different from that of the near extremal limit \cite{Carroll:2009maa}. In fact, in the Eddington-Finkelstein
system, there is a discontinuity between the non-extremal black hole and the extremal one despite the well-defined metrics.
The entropy of extremal black holes still remains as an intriguing one.
Thus, it would be interesting to study the entanglement entropy of a black hole with inner and outer horizons
in order to figure out how the presence of the island influences the entanglement entropy in
extremal black holes.

Starting from a black hole in thermal equilibrium with a heat bath of an arbitrary temperature $\beta$,
one can obtain the area law of the black hole entropy by taking
as $\beta\to \beta_H$ where $\beta_H$ is the inverse temperature of the black hole
\cite{tHooft:1984kcu,Demers:1995dq}. In this respect, we suggest that a black hole with an island
is also assumed to be in thermal equilibrium with a heat bath
of an arbitrary temperature in such a way that
at the interim stage the off-shell generalized entropy $S_{\rm gen}(\kappa)$ can be constructed.
Taking $\kappa\to \kappa_H$, we obtain the generalized entropy
in Eq. \eqref{eq:generalized entropy} as
\begin{equation}
\label{eq:limit}
S_{\rm gen}=\lim_{\kappa \to \kappa_H} S_{\rm gen}(\kappa),
\end{equation}
where $\kappa=2\pi T=2\pi \beta^{-1}$ are off-shell quantities and $\kappa_H$ is the surface gravity at the horizon.
This formula will be used in calculating the entanglement entropy for non-extremal and extremal black holes.
This process is actually trivial for non-extremal black holes but crucial for extremal ones, which will be discussed
in forthcoming sections.
Along the line of Ref. \cite{Hashimoto:2020cas} with the assumption \eqref{eq:limit},
we first revisit the computation of the entanglement entropy
in the non-extremal Reissner-Nordstr\"om black hole with an island and then obtain a slightly different form of the
entanglement entropy from the
previous result in Ref. \cite{Wang:2021woy}. In our case, the entanglement entropy
reduces to that of the Schwarzschild black hole when the electric charge vanishes \cite{Hashimoto:2020cas}.
In the extremal Reissner-Nordstr\"om black hole of our main interest,
we compute the entanglement entropy with and without an island.
To avoid encountering the curvature singularity at the origin in the Reissner-Nordstr\"om black hole,
we repeat the computation of the entanglement entropy for the regular Hayward black hole \cite{Hayward:2005gi}.

The organization of this paper is as follows.
In Sec.~\ref{sec:Basic Strategy and Setups}, we will obtain the generalized entropy from the off-shell generalized entropy
for non-extremal and extremal black holes with and without the island.
In Sec.~\ref{sec:The Entanglement Entropy in Reissner-Nordstr\"om Black holes}, we will
calculate the entanglement entropy of the Reissner-Nordstr\"om black hole.
In Sec.~\ref{sec:The Entanglement Entropy in Hayward Black holes},
the process performed in Sec.~\ref{sec:The Entanglement Entropy in Reissner-Nordstr\"om Black holes}
will be applied to the Hayward black hole.
Finally, conclusion and discussion will be given in Sec.~\ref{sec:conclusion}.

\section{Generalized entropy}
\label{sec:Basic Strategy and Setups}
We study the entanglement entropy of asymptotically flat spherically symmetric
eternal black holes described by a length element
\begin{equation}\label{eq:metric}
	\dd{s}^2 = -f(r)\dd{t}^2 + \frac{1}{f(r)}\dd{r}^2 + r^2\dd{\Omega}^2,
\end{equation}
where $f(r)$ is a metric function which has two horizons $ r_+ $ and $ r_- $ with $ r_+ > r_- $.
The metric in terms of Kruskal coordinates is defined by
\begin{align}\label{eq:Kruskal metric}
	\dd s^2 = -e^{2\rho}\dd{X^+}\dd{X^-} + r^2\dd{\Omega^2},
     \end{align}
where $\kappa X^\pm = \pm e^{\pm\kappa(t \pm r_\ast)}$ and $r_\ast$ is a radial tortoise coordinate
defined by $r_\ast =\int^r (1/f(r))\dd r$,
and $e^{2\rho} = f(r)/(-\kappa^2 X^+ X^-)$.
The Hawking temperature
is calculated as
\begin{equation}\label{eq:9}
	T_H =\beta^{-1}_{H}=\frac{\kappa_H}{2\pi}= \frac{f'(r_+)}{4\pi}
\end{equation}
which is now called the on-shell temperature.

For a finite matter part of the entanglement entropy in curved spacetime,
we consider free massless scalar fields and make an assumption that the distance
between the boundary surfaces is large compared to the scale
of the size of the boundary surfaces so that only the s-wave contributes to the entropy \cite{Hashimoto:2020cas}.
Thus, from the two-dimensional conformal field theory,
the matter part of the entanglement entropy is approximately given as
\begin{equation}\label{eq:ent}
	S_{\rm matter} = \frac{c}{3}\log d(x,y)
\end{equation}
where \( c \) is a central charge for a two-dimensional massless scalar field and \( d(x,y) \) is a distance between
$x$ and $y$. In fact,
the matter part of the entanglement entropy in the logarithmic correction
also includes the proper UV cut-off scale $\epsilon$ \cite{Calabrese:2004eu},
but it will be deleted as an additive constant.

\subsection{Non-extremal black hole}
\subsubsection{Without island}
\begin{figure}
	\subfigure[Without island]{\includegraphics[scale=.70]{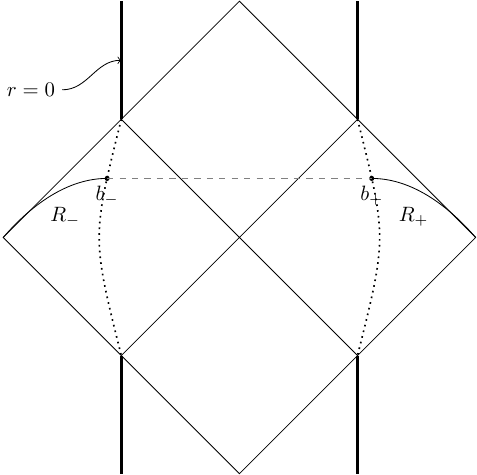}\label{fig:1a}}
	\quad
	\subfigure[With island]{\includegraphics[scale=.70]{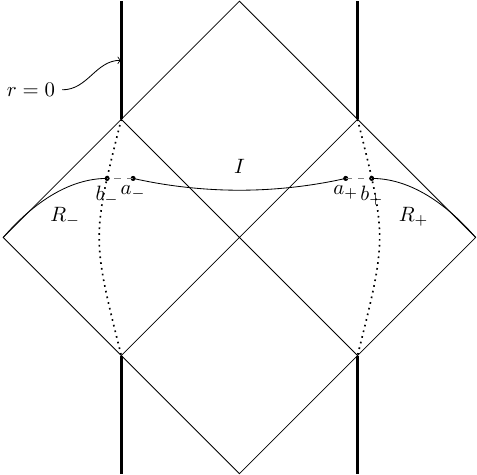}\label{fig:1b}}
	\caption{The Penrose diagrams of an eternal non-extremal black hole are
		presented without an island (left) and with an island (right). The region for the Hawking radiation
		\( R \) is separated into two parts; \( R_+ \) and \( R_- \) located in the right and left wedges, respectively. The boundaries of \( R_+ \) and \( R_- \) are indicated as \( b_+ \) and \( b_- \), respectively. The island extends between
the two wedges and its boundaries are at \( a_+ \) and \( a_- \).
In the Reissner-Nordstr\"om black hole, the curvature is singular at $r=0$ whereas it is finite
there in the Hayward black hole.}
	\label{fig:1}
\end{figure}

The matter part of the generalized entropy
is defined by the entanglement region between
the boundaries $ b_+ = (t_b,b) $ and $ b_- = (-t_b + i\beta / 2, b) $\footnote{
A Schwarzschild coordinate time $ -t + i\beta/2 $ allows us to describe a mirror point located in the left wedge of the Penrose diagram,
which means that $ X^\pm (x_+) = X^\mp(x_-)$ in Kruskal coordinates where $ x_+ = (t,r)  $ in the right wedge and $ x_- = (-t+i\beta/2, r) $ in the left wedge.
This fact can be found from the definition of Kruskal coordinates as
\begin{equation}
	\kappa X^\pm (x_-) = \pm e^{\pm\kappa(-t+\frac{i\beta}{2} \pm r_\ast)} = \pm e^{\pm\kappa(-t+\frac{i\pi}{\kappa} \pm r_\ast)} = \pm e^{i\pi}e^{\pm\kappa(-t \pm r_\ast)} = \mp e^{\mp\kappa(t\mp r_\ast)} = \kappa X^\mp(x_+),\nonumber
\end{equation}
which results in $ X^\pm (x_+) = X^\mp(x_-) $.}
in \cref{fig:1} as
\begin{equation}\label{eq:no island}
	S_{\rm matter} = \frac{c}{3}\log d(b_+,b_-),
\end{equation}
where $d^2(x,y) = -(X^+(x)-X^+(y))(X^-(x)-X^-(y))e^{\rho(x)}e^{\rho(y)}$.
Note that the black hole is in thermal equilibrium with a heat bath of an arbitrary temperature $\beta^{-1}$.
Using Eq. \eqref{eq:limit} and Eq. \eqref{eq:no island}, we have
\begin{equation}\label{eq:entanglement entropy without island}
	S_{\rm gen}=\lim_{\kappa \to \kappa_H} S_{\rm gen}(\kappa)=
 \frac{c}{6}\log\left( \frac{4f(b)}{\kappa^2_H}\cosh^2\kappa_H t_b \right) \approx \frac{c}{3}\kappa_H t_b
\end{equation}
in the late time approximation of $t_b \gg 1/\kappa_H$.
This implies that the entanglement entropy is increasing with respect to time $ t_b $ so that
this entropy becomes much larger than the black hole entropy expected in Ref. \cite{Hawking:1976ra}.
Note that Eq. \eqref{eq:entanglement entropy without island}
in the late time approximation is valid only for non-extremal black holes.

\subsubsection{With island}

The collection of disjoint intervals is defined as
	$R\cup I = R_- \cup I \cup R_+ $,
where the finite boundaries of \( R_+ \) and \( R_- \) are
denoted by \( b_+ \) and \( b_- \), respectively.
The island extends between
the two radiation wedges and its boundaries are at \( a_+ \) and \( a_- \) depicted in \cref{fig:1}.

Let us now introduce a single island in the
calculation of the entanglement entropy.
The matter part with an island whose boundaries located at $ a_+ = (t_a,a) $ and $ a_- = (-t_a + i\beta / 2, a) $
is given by
\begin{equation}\label{eq:with island}
	S_{\rm matter} = \frac{c}{3}\log \frac{d(a_+,a_-)d(b_+,b_-)d(a_+,b_+)d(a_-,b_-)}{ d(a_+,b_-)d(a_-,b_+)}.
\end{equation}
The boundary of \( R \) is assumed to be far away from the outer horizon $ r_+ $,
which makes the s-wave approximation valid, and the island extends vicinity of the outside of the event horizon.
From Eqs. \eqref{eq:generalized entropy}, \eqref{eq:limit} and \eqref{eq:with island},
the generalized entropy for the metric \eqref{eq:metric} is calculated as
\begin{equation}\label{eq:entanglement entropy with island}
	\begin{split}
	S_{\rm gen}=\lim_{\kappa \to \kappa_H}S_{\rm gen}(\kappa) = &\frac{2\pi a^2}{G_N} + \frac{c}{6}\log \left(\frac{16f(a)f(b)}{\kappa_H^4}\cosh^2\kappa_H t_a \cosh^2\kappa_H t_b \right)\\
		&+ \frac{c}{3}\log\left| \frac{\cosh\kappa_H(r_\ast(a)-r_\ast(b))-\cosh\kappa_H(t_a-t_b)}{\cosh\kappa_H(r_\ast(a)-r_\ast(b))+\cosh\kappa_H(t_a+t_b)} \right|.
	\end{split}
\end{equation}
In fact, there is no distinction between the on-shell and the off-shell formalism in non-extremal black holes.

One might wonder in Eqs. \eqref{eq:no island} and \eqref{eq:entanglement entropy without island} why the non-extremal black hole is surrounded by a heat bath of an arbitrary temperature.
Usually non-extremal black holes evaporate in a lower-temperature bath and gain energy in a bath at higher temperatures.
In this work, we assume eternal black holes possessing an arbitrary temperature as an off-shell temperature different from the on-shell temperature defined by the Hawking temperature.
Of course, the heat bath also has the same arbitrary temperature as the black hole temperature.
Therefore, the black hole is in thermal equilibrium with the heat bath and so it does not evaporate.
After calculations, we take the on-shell limit as $ \beta \to \beta_H $, i.e., $ \kappa \to \kappa_H $.
This kind of limiting process in thermal equilibrium has been used in calculating thermodynamic quantities in non-extremal and extremal black holes \cite{tHooft:1984kcu, Demers:1995dq}.

\begin{figure}
	\subfigure[Without island]{\includegraphics[scale=.70]{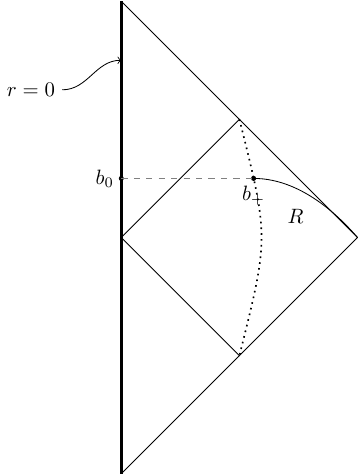}\label{fig:2a}}
	\quad
	\subfigure[With island]{\includegraphics[scale=.70]{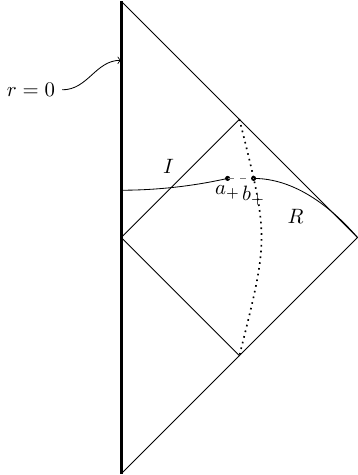}\label{fig:2b}}
	\caption{The Penrose diagrams of an eternal extremal black hole are
		presented without an island (left) and with an island (right). The region for the Hawking radiation
		\( R \) is located in the right wedge. The boundary of \( R \) is denoted by \( b_+ \).
    In (a), a reference point $b_0$ is introduced at $r=0$. In (b), the island extends between
		$r=0$ and \( a_+ \) just outside the horizon.
		In the Reissner-Nordstr\"om black hole, the curvature is singular at $r=0$ whereas it is finite
there in the Hayward black hole.}
	\label{fig:2}
\end{figure}
\subsection{Extremal black hole}
\subsubsection{Without island}

Let us take a spacelike surface which ends at a reference point $ b_0 = (t_0, 0)$ in \cref{fig:2}.
Then, the entanglement entropy can be calculated as
\begin{equation}\label{eq:matter entropy extremal without island}
	S_{\rm matter} = \frac{c}{3}\log d(b_+, b_0),
\end{equation}
and so one can obtain
\begin{align}\label{eq:entanglement entropy extremal without island}
	S_{\rm gen} = \lim_{\kappa \to 0} S_{\rm gen}(\kappa) = \frac{c}{12} \log \left( f(0)f(b)\left[ (r_\ast(b)-r_\ast(0))^2 - (t_b-t_0)^2 \right]^2 \right).
\end{align}
Note that $f(0)$ is singular for the Reissner-Nordstr\"om black hole so that  the generalized entropy becomes infinite.
The infinite entropy in the extremal Reissner-Nordstr\"om black hole is also a consequence of the infinite proper distance of $[b_0,b_+]$.

\subsubsection{With island}

The entanglement-entropy formula for a single interval $R\cup I$ is given by
\begin{equation}\label{eq:single interval with island}
	S_{\rm matter} = \frac{c}{3}\log d(a_+, b_+).
\end{equation}
In \cref{fig:2}, the boundary of \( R \) is assumed to be far away from the horizon $ r_+ = r_- = r_h $.
From Eqs. \eqref{eq:limit} and \eqref{eq:single interval with island},
the off-shell generalized entropy for the classical metric \eqref{eq:metric} for the extremal black hole is calculated as
\begin{equation}\label{eq:single interval entanglement entropy}
	S_{\rm gen}(\kappa) = \frac{ \pi a^2}{G_N} + \frac{c}{12}\log\left( \frac{4f(a)f(b)}{\kappa^4} \right) + \frac{c}{6}\log\left| \cosh\kappa(r_\ast(a)-r_\ast(b)) -\cosh\kappa(t_a - t_b) \right|.
\end{equation}
Taking the limit $\kappa  \to 0 $ in Eq. \eqref{eq:single interval entanglement entropy}, we have a finite value of the
generalized entropy as
\begin{align}\label{eq:exact_ext}
	S_{\rm gen} &=\lim_{\kappa \to 0} S_{\rm gen}(\kappa)\nonumber\\
	&=\frac{\pi a^2}{G_N}+ \frac{c}{12} \log \left(f(a)f(b) \left[ (r_\ast(a)-r_\ast(b))^2 - (t_a-t_b)^2 \right]^2 \right).
\end{align}
Note that the divergences from the second and third terms
in Eq. \eqref{eq:single interval entanglement entropy}
exactly cancel out.

In the next consecutive sections, we will apply the generalized entropy formulas to the Reissner-Nordstr\"om black hole and
the Hayward black hole.

\section{Entanglement Entropy of Reissner-Nordstr\"om Black hole}
\label{sec:The Entanglement Entropy in Reissner-Nordstr\"om Black holes}
The length element of the Reissner-Nordstr\"om black hole is given as
\begin{equation}\label{eq:Reissner Nordstrom Metric}
	\dd{s}^2 = -\left( 1 - \frac{2G_N M}{r} + \frac{G_NQ^2}{r^2} \right) \dd{t}^2 + \left( 1 - \frac{2G_N M}{r} + \frac{G_NQ^2}{r^2} \right)^{-1}\dd{r}^2 + r^2\dd \Omega^2,
\end{equation}
where $ M $ is the black-hole mass and $ Q $ is the electric charge of the black hole.
The non-extremal black hole has two horizons $ r_{\pm} = G_N(M \pm \sqrt{M^2 - Q^2}) $.
If the black hole is extremal: $ M=Q $,
then $r_+=r_-$ and the horizon reduces to $ r_h = G_NM $.
The Reissner-Nordstr\"om black hole has a curvature singularity at $ r=0 $.

\subsection{Non-extremal Reissner-Nordstr\"om black hole}
\label{subsec:RN Non-Extremal Case}
We calculate the entanglement entropy of the non-extremal Reissner-Nordstr\"om black hole described by
the metric function
\begin{equation}\label{eq:RN_metric}
	f(r) = \frac{(r-r_+)(r-r_-)}{r^2}.
\end{equation}
In the absence of the island, the entanglement entropy is nothing but Eq. \eqref{eq:entanglement entropy without island}.

Let us derive the entanglement entropy of the non-extremal Reissner-Nordstr\"om black hole
with an island.
In the late time approximation of $ t_a + t_b \to \infty $ \cite{Hashimoto:2020cas},
the generalized entropy \eqref{eq:entanglement entropy with island} for the metric \eqref{eq:RN_metric} is extremized with respect to $ t_a $ as
\begin{equation}\label{eq:dev_time}
	\pdv{S_{\rm gen}}{t_a} = \frac{c\kappa_H \sinh\kappa_H(t_a-t_b)}
{3[\cosh\kappa_H(t_a-t_b)-\cosh\kappa_H(r_\ast(a)-r_\ast(b))]} = 0,
\end{equation}
which simply yields $ t_a = t_b $.
In Eq. \eqref{eq:dev_time}, the surface gravity $ \kappa_H $ is defined by
\begin{equation}\label{RN surface gravity}
	\kappa_H = \frac{r_+ - r_-}{2r_+^2},
\end{equation}
and the boundaries
of the island and the radiation region in terms of tortoise coordinates are also given by
\begin{align}
  r_{\ast}(a) &= r_+ + \frac{r_+^2}{r_+ - r_-}\log\left( \frac{a-r_+}{r_+} \right),\label{eq:near tortoise}\\
  r_{\ast}(b) &= b + \frac{r_+^2}{r_+ - r_-}\log\left(\frac{b-r_+}{r_+}\right) - \frac{r_-^2}{r_+ - r_-}\log\left(\frac{b-r_-}{r_+}\right). \label{eq:tortoise}
\end{align}

Next, the derivative of the generalized entropy \eqref{eq:entanglement entropy with island}
with respect to $ a $ is obtained as
\begin{align}\label{eq:dev_rad}
	\pdv{S_{\rm gen}}{a}
			   &= \frac{4\pi a}{G_N} + \frac{c}{6f(a)}\left[ f'(a) - 2\kappa_H - 4\kappa_H e^{\kappa_H (r_\ast(a) - r_\ast(b))} \right] = 0
\end{align}
by using the fact that $ t_a = t_b $, $ t_a + t_b \to \infty $, and $ a \sim r_+ $.
From Eq. \eqref{eq:dev_rad}, the boundary of the island can approximately be determined as
\begin{align}\label{eq:island_nonext}
	a =  r_+ + \frac{(cG_N)^2}{144\pi^2 r_+^3}e^{2\kappa_H(r_+-r_\ast(b))} + \mathcal{O}(G_N^3).
\end{align}

From Eq. \eqref{eq:generalized entropy}, the entanglement entropy of the non-extremal Reissner-Nordstr\"om black hole
with an island
 is obtained as
\begin{align}\label{eq:ent_nonext}
	S_{\rm I} &= \frac{2\pi r_+^2}{G_N} + \frac{c}{6}\left[\log\left( \frac{16r_+^{6}(b-r_+)^2(b-r_-)}{b^2(r_+ - r_-)^3}\right)-\frac{r_-^2}{r_+^2}\log(\frac{b-r_-}{r_+})+\frac{r_+ - r_-}{r_+^2}(b-r_+)\right].
\end{align}
The expression of the entanglement entropy \eqref{eq:ent_nonext} looks different from that in Ref. \cite{Wang:2021woy}
in the sense that the Schwarzschild limit appears to be different. For $ r_- = 0 $, Eq. \eqref{eq:ent_nonext}
exactly reduces to the entanglement entropy of the Hawking radiation in the Schwarzschild black hole \cite{Hashimoto:2020cas},
\begin{equation}\label{Schwarzschild Entropy}
	S_{\rm I} = \frac{2\pi r_h^2}{G_N}+ \frac{c}{6}\left[ \log\left( \frac{16r_h^3(b-r_h)^2}{b} \right) + \frac{b-r_h}{r_h} \right].
\end{equation}

In the non-extremal Reissner-Nordstr\"om black hole,
the Page time can be estimated
at which the growing entanglement
entropy \eqref{eq:entanglement entropy without island} changes into the constant entanglement entropy \eqref{eq:ent_nonext}.
The Page time is the same form as that of the Schwarzschild black hole in Ref. \cite{Hashimoto:2020cas},
which is almost universal,
\begin{equation}\label{eq:Page time}
	t_{\rm Page} = \frac{6\pi r_+^2}{\kappa_H cG_N} = \frac{3S_{\rm BH}}{\pi c T_{\rm H}},
\end{equation}
where $S_{\rm BH} = (\pi r_+^2)/G_N$ is the Bekenstein-Hawking entropy. The Page time is getting longer when the black-hole mass is close to the electric charge.
\subsection{Extremal Reissner-Nordstr\"om black hole}
Let us calculate the entanglement entropy of the extremal Reissner-Nordstr\"om black hole described by the metric function as
\begin{equation}\label{eq:RN_ext_metric}
	f(r) = \frac{(r-r_h)^2}{r^2}.
\end{equation}

We investigate the entanglement entropy of the extremal Reissner-Nordstr\"om black hole with an island.
Extremizing the generalized entropy \eqref{eq:exact_ext} for the metric \eqref{eq:RN_ext_metric}
with respect to $ t_a $ gives
\begin{equation}\label{eq:dev_time_ext}
	\pdv{S_{\rm gen}}{t_a} = \frac{c (t_a-t_b)}{3[(t_a-t_b)^2-(r_\ast(a)-r_\ast(b))^2]} = 0,
\end{equation}
where $r_{\ast}(a)$ and $r_{\ast}(b)$ are obtained through Eq. \eqref{eq:RN_ext_metric} as
\begin{align}
	r_{\ast}(a) &= r_h - \frac{r_h^2}{a-r_h},\label{eq:near_tortoise_ext}\\
	r_{\ast}(b) &= b - \frac{r_h^2}{b-r_h} + 2r_h\log\left( \frac{b - r_h}{r_h} \right)\label{tortoise ext}.
\end{align}
In this case, without resort to the late time approximation of $ t_a+t_b \to \infty $, one can simply get $ t_a = t_b $.

Next, the derivative of the generalized entropy \eqref{eq:exact_ext} with respect to $ a $ is found to be
\begin{equation}\label{eq:dev_rad_ext}
	\pdv{S_{\rm gen}}{a} = \frac{2\pi a}{G_N} + \frac{c}{12f(a)}\left( f'(a) + \frac{4}{r_\ast(a) - r_\ast(b)} \right) = 0
\end{equation}
within the approximation $a\sim r_+$.
Hence, the boundary of the island can be determined as
\begin{align}\label{eq:island_ext}
	a = r_h + \frac{cG_N}{12\pi r_h} + \frac{(cG_N)^2(7r_h - 2r_\ast(b))}{144\pi^2r_h^4}+ \mathcal{O}(G_N^3).
\end{align}
From Eq. \eqref{eq:generalized entropy},
the entanglement entropy of the extremal Reissner-Nordstr\"om black hole with an island is finally obtained as
\begin{align}\label{eq:ent_ext}
	S_{\rm I} =\frac{\pi r_h^2}{G_N} + \frac{c}{6} \log\left(\frac{12\pi r_h^4(b-r_h)}{cG_Nb}\right).
\end{align}
Note that the entanglement entropy \eqref{eq:ent_ext} cannot be obtained from
the continuous extremal limit of the non-extremal entanglement entropy \eqref{eq:ent_nonext} because
the extremal case and the extremal limit are in essence different \cite{Carroll:2009maa}.

In \cref{fig:2},
the island hits the boundary of the island and $f(0)$ is also divergent so that
the semi-classical approximation in that region appears to be invalid.
In this regard, we would repeat the computation of the entanglement entropy for a regular black hole
in the next section.

\section{Entanglement Entropy of Hayward Black hole}
\label{sec:The Entanglement Entropy in Hayward Black holes}
In this section, we study the entanglement entropy of
the Hayward black hole described by the metric \cite{Hayward:2005gi}
\begin{equation}\label{Hayward Metric}
	\dd{s}^2 = -\left( 1 - \frac{2G_NMr^2}{r^3 + 2G_NM\ell^2} \right) \dd{t}^2 + \left( 1 - \frac{2G_NMr^2}{r^3 + 2G_NM\ell^2} \right)^{-1}\dd{r}^2 + r^2\dd \Omega^2,
\end{equation}
where $ M $ is the black-hole mass and $ \ell $ is the Hubble length.
The metric is asymptotically flat such that it approaches the Schwarzschild metric as
\begin{equation}\label{Hayward Far metric}
	f(r) \sim 1 - \frac{2G_NM}{r} \qas r\to \infty.
\end{equation}
Note that the metric behaves like the de Sitter space near the origin $ r=0 $ as
\begin{equation}\label{Hayward Origin Metric}
	f(r) \sim 1 - \frac{r^2}{\ell^2} \qas r \to 0,
\end{equation}
where the curvature is finite at the origin.
Unlike the Reissner-Nordstr\"om black hole, the Hayward black hole is regular everywhere.

\subsection{Non-extremal Hayward black hole}
In Eq. \eqref{Hayward Metric}, the metric function of the non-extremal Hayward black hole is neatly
written as
\begin{equation}\label{eq:hay_metric}
	f(r) = \frac{(r-r_+)(r-r_-)(r-r_0)}{r^3-r_+r_-r_0},
\end{equation}
where $ r_+ $ and $ r_- $ are two positive roots satisfying
\begin{equation}\label{eq:hay_equ}
	r^3 - 2G_NMr^2 + 2 G_NM \ell^2 = 0,
\end{equation}
and $ r_0 = 2G_NM - (r_+ + r_-) $ which is negative.
If $ M > (3\sqrt{3}/4G_N)\ell $, the black hole has two horizons: $ r_+ $ and $ r_- $
where $r_+ > r_-$.
For $ r_- = r_0 = 0 $, the Hayward black hole reduces to the Schwarzschild black hole.
In the absence of the island,
the entanglement entropy is the generalized entropy \eqref{eq:entanglement entropy without island}.

Let us calculate the entanglement entropy of the non-extremal Hayward black hole with an island.
Extremizing condition of the generalized entropy \eqref{eq:entanglement entropy with island}
for the metric \eqref{eq:hay_metric} with respect to $ t_a $
is given as
\begin{equation}\label{eq:dev_time:Hayward}
	\pdv{S_{\rm gen}}{t_a} = \frac{c\kappa_H \sinh\kappa_H(t_a-t_b)}{3[\cosh\kappa_H(t_a-t_b)-\cosh\kappa_H(r_\ast(a)-r_\ast(b))]} = 0
\end{equation}
in the late time approximation of $ t_a + t_b \to \infty $, and thus, one can find $ t_a = t_b $.
In Eq. \eqref{eq:dev_time:Hayward}, the surface gravity is defined by
\begin{equation}\label{Hayward surface gravity}
	\kappa_H = \frac{(r_+ - r_-)(r_+-r_0)}{2(r_+^3 - r_+r_-r_0)},
\end{equation}
and the boundaries of the island and the radiation region in terms of the tortoise coordinates are written as
\begin{align}
  r_{\ast}(a) &= r_+ +  \frac{r_+^3 - r_+r_-r_0}{(r_+ - r_-)(r_+ - r_0)}\log\left( \frac{a-r_+}{r_+} \right),\label{eq:hay near tortoise}\\
  r_{\ast}(b) &= b + \frac{r_+^3 - r_+r_-r_0}{(r_+ - r_-)(r_+ - r_0)}\log\left(\frac{b-r_+}{r_+}\right) - \frac{r_-^3 - r_+r_-r_0}{(r_+ - r_-)(r_- - r_0)}\log\left(\frac{b-r_-}{r_+}\right) \nonumber\\
  &\qquad + \frac{r_0^3 - r_+r_-r_0}{(r_+ - r_0)(r_- - r_0)}\log\left(\frac{b-r_0}{r_+}\right). \label{eq:hay tortoise}
\end{align}

Next, the extremum condition with respect to $a$ is given as
\begin{align}\label{eq:hay_dev_rad_approx}
	\pdv{S_{\rm gen}}{a} &= \frac{4\pi r_+}{G_N} -\frac{cr_+}{r_+^2 -r_-r_0} + \frac{c}{3}\left( \frac{1}{r_+-r_-} + \frac{1}{r_+-r_0}-\frac{1}{\sqrt{r_+(a-r_+)}}e^{\kappa_H(r_+-r_{\ast}(b))} \right) = 0,
\end{align}
where $ t_a = t_b $ was used with the late time approximation $ t_a + t_b \to \infty $.
From Eq. \eqref{eq:hay_dev_rad_approx}, the boundary of the island is determined as
\begin{align}\label{eq:hay_island_nonext}
	a &\simeq r_+ + \frac{(cG_N)^2}{144\pi^2 r_+^3}e^{2\kappa_H(r_+-r_\ast(b))} + \mathcal{O}(G_N^3).
\end{align}
Hence, the entanglement entropy of the non-extremal Hayward black hole with an island is obtained as
\begin{align}\label{eq:hay_ent_nonext}
	S_{\rm I} &= \frac{2\pi r_+^2}{G_N} + \frac{c}{6}\left[\log\left( \frac{16(r_+^3-r_+r_-r_0)^3(b-r_+)^2(b-r_-)(b-r_0)}{(b^3-r_+r_-r_0)(r_+-r_-)^3(r_+-r_0)^3} \right)\right.\\
	& - \frac{r_-^3 - r_+r_-r_0}{r_+^3 - r_+r_-r_0}\log(\frac{b-r_-}{r_+})+\frac{r_0^3 - r_+r_-r_0}{r_+^3 - r_+r_-r_0}\log(\frac{b-r_0}{r_+}) \nonumber \\
	&\left.+\frac{(r_+ - r_-)(r_+ - r_0)}{r_+^3 - r_+r_-r_0}(b-r_+)\right] \nonumber.
\end{align}
For $ r_-= r_0 = 0 $,
Eq. \eqref{eq:hay_ent_nonext} exactly reduces to the entanglement entropy of the Schwarzschild black hole \eqref{Schwarzschild Entropy}
in Ref. \cite{Hashimoto:2020cas}.
In addition, the Page time can be obtained
at which the growing entanglement
entropy \eqref{eq:entanglement entropy without island} changes into the constant entanglement entropy \eqref{eq:hay_ent_nonext}
in the leading order. The Page time is the same form as Eq. \eqref{eq:Page time}.

\subsection{Extremal Hayward black hole}
For $ M = (3\sqrt{3}/4G_N)\ell $ and $r_0 = (-\sqrt{3}/2)\ell$,
the horizons are degenerate as
$ r_+=r_-=r_h = \sqrt{3}\ell $, and so the metric function for the extremal Hayward black hole is given as
\begin{equation}\label{eq:hay_RN_ext_metric}
	f(r) = \frac{(r-r_h)^2(2r+r_h)}{2r^3 + r_h^3}
\end{equation}
where $ r_h = (4/3)G_NM$.

By using Eq. \eqref{eq:entanglement entropy extremal without island},
the entanglement entropy of the extremal Hayward black hole without an island is calculated as
\begin{align}\label{eq:entanglement entropy extremal without island Hayward}
	S=S_{\rm gen} \approx \frac{c}{3}\log t_b,
\end{align}
where we ignored the other finite terms in late times and used the fact that $f(0)=1$
in Eq. \eqref{eq:hay_RN_ext_metric}.

We now compute the entanglement entropy of the
extremal Hayward black hole with an island by using Eq. \eqref{eq:exact_ext}.
The extremization condition of the entanglement entropy \eqref{eq:exact_ext} with respect to $ t_a $ is
\begin{equation}\label{eq:Hayward dev_time_ext}
	\pdv{S_{\rm gen}}{t_a} = \frac{c (t_a-t_b)}{3[(t_a-t_b)^2-(r_\ast(a)-r_\ast(b))^2]} = 0,
\end{equation}
and thus, $ t_a = t_b $ is found without the late time approximation.
In Eq. \eqref{eq:Hayward dev_time_ext}, the tortoise coordinates for the boundaries are given as
\begin{align}
	r_{\ast}(a) &= r_h - \frac{r_h^2}{a-r_h},\label{eq:hay_near_tortoise_ext}\\
	r_{\ast}(b) &= b - \frac{r_h^2}{b-r_h} + \frac{4r_h}{3}\log\left( \frac{2(b-r_h)}{r_h} \right) + \frac{r_h}{6}\log\left( \frac{2b + r_h}{r_h} \right)\label{eq:hay tortoise ext}.
\end{align}

Next, the derivative of the generalized entropy \eqref{eq:exact_ext} with respect to $ a $ is found to be
\begin{align}\label{eq:hay_dev_rad_ext_approx}
	\pdv{S_{\rm gen}}{a} &= \frac{2\pi r_h}{G_N} - \frac{c}{6(a-r_h)} -\frac{8c}{9r_h}+\frac{c}{3r_h^2}r_\ast(b)=0.
\end{align}
If the island boundary is assumed to be very close to the horizon, $a\sim r_+$,
then the boundary of the island is approximately determined as
\begin{align}\label{eq:hay_island_ext}
	a &\simeq r_h + \frac{cG_N}{12\pi r_h} + \frac{(cG_N)^2(8r_h - 3r_\ast(b))}{216\pi^2r_h^4} + \mathcal{O}(G_N^3).
\end{align}
Thus, from Eq. \eqref{eq:generalized entropy},
the entanglement entropy of the extremal Hayward black hole with an island is finally obtained as
\begin{align}\label{eq:hay_ent_ext}
	S_{\rm I} &= \frac{\pi r_h^2}{G_N} + \frac{c}{12} \log\left(\frac{144\pi^2r_h^8(b-r_h)^2(2b+r_h)}{(cG_N)^2(2b^3+r_h^3)}\right).
\end{align}
The extremal entanglement entropy \eqref{eq:hay_ent_ext} cannot be obtained by taking the extremal limit of Eq. \eqref{eq:hay_ent_nonext}.

Now, the Page time can be obtained
at which the growing entanglement
entropy \eqref{eq:entanglement entropy extremal without island Hayward} changes into the constant entanglement entropy \eqref{eq:hay_ent_ext}
as
\begin{equation}\label{extremal Page time}
	t_{\rm Page} = \left(\frac{144\pi^2r_h^8(b-r_h)^2(2b+r_h)}{(cG_N)^2(2b^3+r_h^3)}\right)^{1/4}e^{\frac{3}{c}S_{\rm BH}},
\end{equation}
in which we take into account up to the sub-leading logarithmic term in Eq. \eqref{eq:hay_ent_ext}, and
$S_{\rm BH} = (\pi r_+^2)/G_N$ is the Bekenstein-Hawking entropy.
In the large black hole but $b \gg r_h$, the Page time in the extremal black hole \eqref{extremal Page time} appears to be longer than Eq. \eqref{eq:Page time}
in the non-extremal black hole.

Let us explicate some special properties of the Hayward black hole in calculations of the entanglement entropy in the island proposal.
In general relativity, the Schwarzschild black hole is geodesically incomplete in the presence of a curvature singularity responsible for breakdown of classical and quantum predictabilities.
Fortunately, the curvature singularity is protected by the event horizon so that internal structures of the black hole are of no relevance to outside observers.
A curvature singularity also appears in the Reissner-Nordstr\"om black hole.
In the same manner, the curvature singularity is enclosed by the event horizon.
But the latter case provides an extremal limit in contrast to the Schwarzschild black hole case.

In the island prescription, a curvature singularity inside the horizon may affect the calculation of the entanglement entropy of extremal black holes.
Let us now mention what it happens in the extremal Reissner-Nordstr\"om black hole.
In the calculations of the generalized entropy before the Page time in \cref{fig:2a}, the entanglement entropy in the radiation region was not well-defined because of the curvature singularity at $ b_0 = (t_0,0) $.
Explicitly, as seen from Eq. \eqref{eq:entanglement entropy extremal without island}, the lapse function is divergent, i.e., $ f(0) = \infty $, which makes the generalized entropy divergent whereas $ f(0) $ is finite in the Hayward black hole.
Recall that a regularization of curvature singularity was unnecessary in the non-extremal case in \cref{fig:1a} because the boundaries of the entanglement wedges never touch the curvature singularity.
Usual islands or their boundaries were always positioned in a finite curvature region.
On the other hand, after the Page time, the internal structure of the extremal Reissner-Nordstr\"om black hole also plays an important role.
In \cref{fig:2b}, the left boundary of the island necessarily ends at the origin; in other words, the timelike curvature singularity encounters the boundary of the spacelike island inside the horizon.
In contrast to usual islands in a finite curvature region, the island in the extremal Reissner-Nordstr\"om black hole encounters the curvature singularity.
After the Page time, we have to consider the island contribution to the entanglement entropy for radiation but we do not understand how to handle the entanglement entropy formula when interacts between the particles on the left boundary of the island and the infinite curvature.

Fortunately, the Hayward black hole is asymptotically flat and everywhere regular. Especially, it has a de Sitter core: $ f(r) \approx 1-r^2/\ell^2 $.
Moreover, it has two horizons like the Reissner-Nordstr\"om black hole.
In order to study the entanglement entropy in the extremal case of a certain black hole with two horizons, it would be nice to consider a regular black hole such as the Hayward black hole.
Before the Page time, in the extremal Hayward black hole, we can get the finite entanglement entropy \eqref{eq:entanglement entropy extremal without island Hayward} because the lapse function is finite, that is, $ f(0) = 1 $.
This finite result is different from the divergent entanglement entropy \eqref{eq:entanglement entropy extremal without island} of the extremal Reissner-Nordstr\"om black hole with a singularity. After the Page time, we also obtained the entanglement entropy \eqref{eq:hay_ent_ext} without encountering complicated issue on interactions between the particles on the boundary of the island and the curvature singularity. These two features before and after the Page times are certainly advantages of the Hayward black hole.

At first sight, one might be tempting to regard the geometry of the extremal Reissner-Nordstr\"om black hole simply as the $ AdS_2 \cross S^2 $ so that the spacetime might be regular like the Jackiw-Teitelboim model in Ref. \cite{Almheiri:2019yqk}.
However, the extremal Reissner-Nordstr\"om black hole reduces $ AdS_2 \cross S^2 $ only near the horizon.
From the full geometrical view-point, the curvature singularity still exists inside the horizon and affects calculations of the entanglement entropy of the extremal black hole in the island proposal.
The whole geometry of the extremal Reissner-Nordström black hole is different from that of Jackiw-Teitelboim model in spite of providing almost the same entropy results.
Consequently, in the island proposal, the Hayward black hole is appropriate to study the entanglement entropy of an extremal black hole.

\section{Conclusion and Discussion}
\label{sec:conclusion}
In conclusion, we investigated the entanglement entropy of the Reissner-Nordstr\"om black hole and the Hayward black hole
for the non-extremal and extremal cases.
In our calculations,
the eternal black holes with an island were assumed to be {\it a priori} in thermal equilibrium with a heat bath of
an arbitrary temperature.
Taking the on-shell condition, we obtained the generalized entropy and then found the entanglement entropy for each case.
In particular, the island rule provided the well-defined fine-grained entropy even in the extremal black hole
as long as the curvature is finite.

In connection with a heat bath of an arbitrary temperature,
the present off-shell formulation of the entropy actually rests upon previous investigations
calculating the statistical entropy of a scalar field on a classical black-hole background \cite{tHooft:1984kcu,Demers:1995dq}.
The thermal gas is assumed to be in thermal equilibrium with a thermal bath of an arbitrary temperature.
The classical geometry and the thermal bath are treated independently because the thermal effect is intrinsically
of no relevance
to the classical geometry. After calculations of
the statistical entropy for the arbitrary temperature, the temperature is finally set by the Hawking temperature,
which results in the area law of the entropy.

Now we make a couple of comments.
First, one of the distinctions of the Reissner–Nordstr\"om black hole from the
Schwarzschild black hole is the appearance of the inner horizon.
The Reissner–Nordstr\"om black hole appears to be unstable under
small perturbations due to the mass inflation at the inner horizon \cite{Poisson:1990eh};
however, we did not consider the mass inflation which is also responsible for the strong cosmic censorship.
In our work, the eternal black holes were just assumed in order to figure out how the island rule can be applied to asymptotically flat
black holes with inner and outer horizons.
Second, in our calculations for the extremal Reissner–Nordstr\"om black hole with an island,
the left-boundary of the island reached the singularity, which would be beyond the semi-classical approximations
in that region. Of course, this issue does not appear
in the Hayward black hole regular at the origin, but it needs some further investigations.
Finally, the island formula was first derived from black holes in anti-de Sitter spacetime coupled to a non-gravitational bath
based on assumptions about factorization of the Hilbert space.
Therefore, the Page curve in asymptotically flat spacetime
may fail in the sense that
information is always available outside a black hole in flat space even before the Page time \cite{Laddha:2020kvp,Raju:2020smc}.
Even weak effects from a gravitational bath can provide different answers from local theories \cite{Geng:2020fxl}.
It means that concrete theoretical investigations are needed in such a way that
the island rule can be well applied to asymptotically flat spacetimes.

\acknowledgments
We would like to thank Hwajin Eom, Jeongwon Ho, Sojeong Jung, and Sang-Heon Yi for helpful and stimulating discussions.
We also wish to thank Suvrat Raju for helpful comments on heat baths.
This research was supported by Basic Science Research Program through the National Research Foundation of Korea(NRF) funded by
the Korea government(MSIP) (Grant No.~2017R1A2B2006159) and
the Ministry of Education through the Center for Quantum Spacetime (CQUeST) of Sogang University (NRF-2020R1A6A1A03047877).


\bibliographystyle{JHEP}       

\bibliography{references.bib}
\end{document}